# Unseen Power of Information Assurance over Information Security

Mouanda, guy

*Abstract*

*Information systems and data are necessary resources for several companies and individuals, but they likewise encounter numerous risks and dangers that can threaten their protection and value. Information security and information assurance are two connected expressions of protecting the confidentiality, integrity, and availability of information systems and data. Information security relates to the processes and methods that block unlawful entry, reform, or exposure of data; in contrast, information assurance covers the expansive aspirations of ensuring that data is responsible, consistent, and flexible. This paper leads the primary models, rules, and challenges of information security and information assurance, examines some of the top methods, principles, and guidelines that can aid in reaching them, and then investigates the modification in prominence for information security from being obscured in the information technology field to a liability pretending to be in the middle resolving all technology breaches around the world. This paper weighs the various controls and how information assurance can be used to spotlight security problems by focusing on human resource assets and technology. Finally, it demonstrates how information assurance must be considered above others' technology pretending to secure the information.*

***Keywords***: *Risk Assessment, Securities Models, Authentication, Access Control, Network Security, Cryptography, Policies, Software Security. Information assurance, Information security*

## 1-Introduction

Information is one of the highest worthful possessions in today's world, as it permits-verdict making, exchange, relationships, invention, and knowledge. Information structures are the hardware, software, and network modules that collect, process, diffuse, and display information. On the contrary, information and information methods are also subject to many risks and dangers that may affect their defensibility and quality. For example, malevolent actors such as hackers, villains, saboteurs, or insiders may try to snipe, change, or annihilate data, or interrupt or disallow its accessibility. Information security and information assurance are two interconnected subjects that point to defending the confidentiality, integrity, and availability of information and its fundamental approaches

Information security and information assurance are confronting and multifaceted fields, as they entail a full and multidisciplinary methodology that judges the specialized, organizational, human being, and conservational considerations that touch data and organizations. In addition, information security and information assurance are dynamic and rising foundation, as they desire to control the shifting needs, requirements, and prospects of data consumers and suppliers, and the developing technologies, trends, and risks that outline the information scenery. Consequently, information security and information assurance demand persistent knowledge, renovating, and adjusting the awareness, proficiencies, and preparations that facilitate data and systems security.

This document launches the central thoughts, standards, and challenges of cyber-security and information assurance and some of the best practices and values that can facilitate reaching them. Also, encloses areas such as security models, policies, mechanisms, cryptography, authentication, access control, network security, web security, software security, database security, cloud security, mobile security, risk assessment, incident response, disaster recovery, and auditing. The document is indented for students, researchers, practitioners, and everyone studying additional security and information assurance.

## 2. Background

Information assurance and security are disturbed by the proposal, application, and assessment of policies, values, procedures, and devices that aim to stop, identify, and reply to cyber threats, as well as to improve from and study cyber events. IAS includes numerous areas, such as cryptography, network security, system security, application security, web security, cloud security, mobile security, IoT security, biometric security, digital forensics, cyber intelligence, cyber risk management, cyber law, cyber ethics, cyber governance, cyber diplomacy, and cyber culture. Information assurance and security likewise encompass dissimilar performers, such as administrations, industries, academia, civil society, and individuals, who have dissimilar parts, accountabilities, and interests in confirming the security and dependability of the data and communication substructures and facilities. Information assurance and security is an evolving and dynamic field that faces constant challenges and opportunities due to cyberspace's fast-changing nature and complexity of cyber threats.

Information assurance and security needs an all-inclusive and interdisciplinary method that reflects the human, social, structural, and conservational influences that touch the security and flexibility of the data and communication approaches and networks. There are three basics corrective types to protect information and data: technology, operations awareness, training, and education. [1] Furthermore, IAS desires to poise the trade-offs and battles that may occur among diverse security purposes, such as availability, confidentiality, integrity, accountability, and usability, as well as amid security and additional standards, such as privacy, freedom, modernization, and growth. Information assurance and security similarly wish fit the diverse backgrounds and situations that may require several stages and categories of security, such as personal, corporate, national, provincial, and international.

## 3. Philosophy of Securities

The conventional task of information security is to defend perceptive information from illicit behaviors, plus check up, revision, recording, and any trouble or damage. The goal is to ensure the safety and privacy of critical data such as customer account details, financial data, or intellectual property. This mission is feared to be accomplished. IT departments of some large corporate organizations are becoming a luxurious identity lacking strategies or frameworks to defend the system or attack unwanted issues, resulting in data loss incidents where millions of customers' records have been exposed. Information security weakness became a buzz in national and international media. A board that neglects information assurance in their company can pay a price of damages. Why has information security become a priority in our society and gotten more attention from the government and private sectors? There are multiple reasons; maybe because the world is reduced in seconds digitally, and the lack of borders creates multiple damages at all levels of society.

Information has always been valuable to someone; he is the living asset of a company organization and is required to keep it safe. The question of how to secure information is no longer reserved for the IT professional but involves everyone viewing all characteristics of security from all viewpoints. There are requirements to be knowledge of the information security risks and weaknesses and then explaining, and adopting a complete solution. While it would be vast to think there was an easy tool to patch up all security troubles, the truth is that there is not. The reasonable approach needs to put on table the information assurance because it involves people and processes together. Technology itself is not enough.

Information security has been developed around the world. People focus on damage caused by a lack of security appropriate to secure the system; others complain about an insufficient tool capable of securing the system and finally, most complaints come from the sum of money they expend to be safe digitally, although technology is a main portion of any business security approach, it is not plentiful. Security needs to be contemplated in every single IT endeavor, but it must be tailored to the obligations and wishes of the corporation. Information security has given methods to additional complete information assurance, focusing more on several strategic, operational, and tactical controls.

The traditional role of information security is to examine the moral, social, political, and legal effects of protecting data and communication techniques from risks and incidents or attacks.

- Moral: the evaluation of the moral values and theories that drive the performance and choices of cyber security professionals and consumers
- Social: the assessment of the influence of cyber security on human connections, background, personality, and society

- Political: the assessment of the dominant changing aspects, interests, and disagreements that manipulate cyber security plans and methods at local, national, and international stages
- Legal: the assessment of the regulations and rules that manage cyber security privileges and obligations, as well as the challenges and chances for legal execution and agreement

## 4. Misconception of Information Security

Information security is a set of security techniques and tools that approximately defend tricky enterprise data from mishandling, unconstitutional access, interruption, or damage. His responsibility is to supervise organization systems for security violations and examine then when they arise. The 3 pillars of information security are what we well-known CIA triad. Figure 1.

- Confidentiality
- Integrity
- Availability

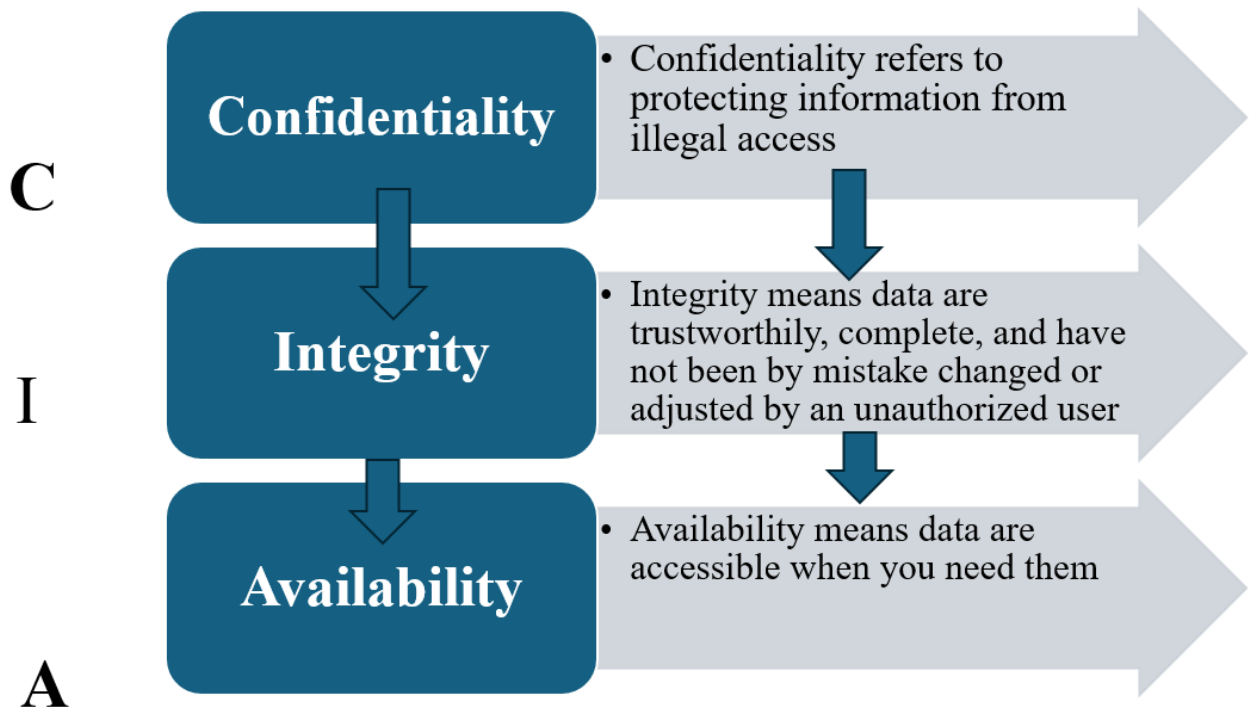

CIA Triad: The 3 pillars of Information security

Guy Mouanda

Figure 1 CIA Pillars of information security

Nowadays, information security is no longer considered a sufficient option to fight against insecurity. It attempts to pressure individuals to think that encryption or anti-virus, firewall, and backup are sufficient to secure the system. People believe security is about technological point solutions that are beyond their reach hidden in the deepest recesses of the IT department**. [2]** This fails to see the intention of the substantial desire for security today. – We need to point out numerous solutions other than just the technology and involve every single member of staff in acquiring a complete solution. Technology helps and can solve a problem, but it is the people and the procedures surroundings it that ensures the information is safe.

The traditional role of information security was reserved for the strict staff of the IT department or to the CIO. Nowadays, with control of government and lawmaking, the answer is everyone. Information security must begin with the CEO and the board and external beyond the business edge to include suppliers, partners, consultants, contractors, and in some situations, customers, all of whom must be part of the community securing information. Whether someone is responsible for data or not should never occur. [2].

## 5. Information Assurance

Today, many people are still confused between information assurance and information security or information technology.NIST SP 800-12 Rev. 1  [3]. Information assurance in CNSSI 4009 is defined as measures that protect and defend information and information systems by ensuring their availability, integrity, authentication, confidentiality, and non-repudiation. These measures include providing for restoration of information systems by incorporating protection, detection, and reaction capabilities. For NIST SP 800-59 under information security 44 U.S.C., Sec. 3542 (b)(1) [4]. The term 'information security' means protecting information and information systems from unauthorized access, use, disclosure, disruption, modification, or destruction to provide integrity, confidentiality, and availability. Computer systems, software, programming languages and data, as well as information processing and storage, all fall under the umbrella of information technology. It is part of information and communication technology [5].

   1-Pillars of Information assurance

Information assurance (IA) is fundamentally defending data systems and is frequently linked with the following five pillars. Fig 2:

- Integrity
- Availability
- Authentication
- Confidentiality
- Non- repudiation

A security audit will outline the primary information security errors in your company and demonstrate you where it is and isn't satisfying the principles it has set out to reach. For businesses handling susceptible and private data, security audits are necessary to creating risk estimation and improvement plan

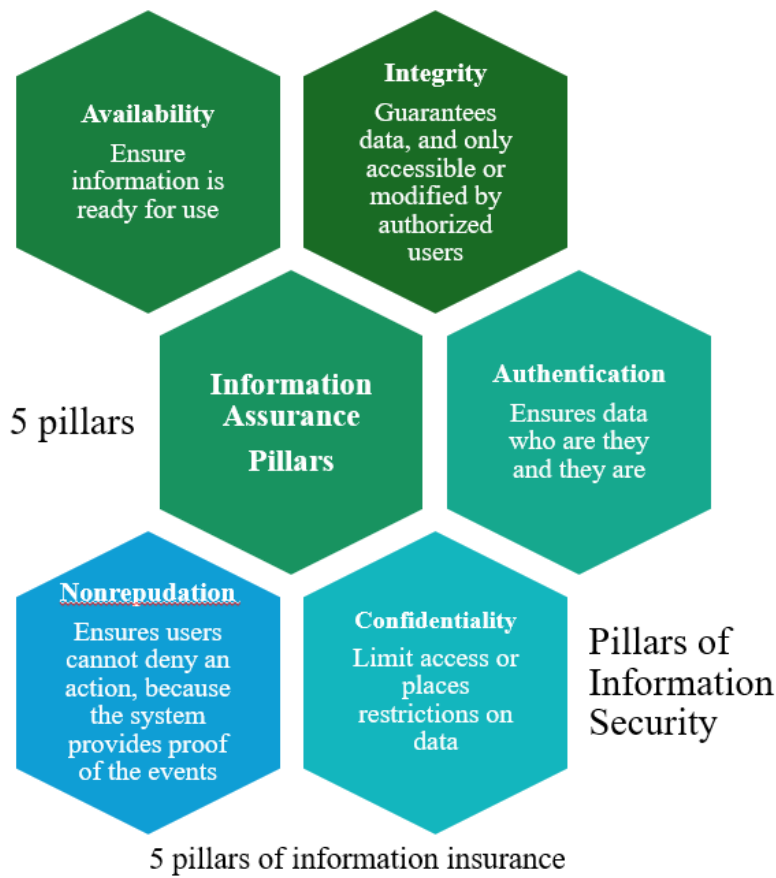

Fig 2: Explains a 5 pillars of information assurance

Those five pillars can be concerned in several circumstances, depending on the responsiveness of your organization's information or information systems. Information assurance is very important to controlling the confidentiality, integrity, and availability of information assets. To guarantee the ethical administration of information, it is vital to perform quality information assurance, regardless of the storage method employed for information assets. [6] This embraces the measures installed to uphold the confidentiality of the information. Confidentiality of information supports privacy of sensitive information that recognize users, such as their name, address, birth date, social security number, credit card numbers, bank information, and medical health data. [7] Integrity in information assurance grants a structure of truth and strength that facilitates the information to be utilized for professional and personal reasons without being unacceptable and biased. [6, 7] When data is without integrity, there is a risk that the information is imprecise and prejudiced. Bias in data creates treachery. When information assurance symbolizes confidentiality, integrity, and availability, it sides with the proposals for information security practices and procedures recommended by the National Security Agency and the U.S. Department of Homeland Security. [7] Authentication refers to the certification and authority within an information structure. Integrity refers to the security of information from illicit modification. Availability refers to the protection of data to be customized. Confidentiality refers to the measure that protects against who can access the data. Non-repudiation refers to the honesty of the data being accurate to its source.

Information assurance is the process of guarantying that the primary pillars of information system are met all over its lifecycle by utilizing a combination of technologies and human intelligence. These 5 pillars -integrity, availability, authentication confidentiality, and non repudiation -are not independent. Interaction between them can be problematic. For instance, availability establishes conflicts with at least three of the others, four pillars: confidentiality, integrity, and authentication [7].

*2- Techniques*

There are two mega techniques with information assurance: audit and risk assessment.

- Audit. A security audit is an all-inclusive assessment of information systems; naturally, this assessment checks a company system's security best practices, externally established standards, and/or federal regulations. A security audit will outline the primary information security errors in your company and demonstrate you where it is and isn't satisfying the principles it has set out to reach. For businesses handling susceptible and private data, security audits are necessary to creating risk estimation and improvement plan.[8]

- Risk assessment. Risk assessment is a precise exploration to detect IT security vulnerabilities and risks. is to guarantee that required security controls are included in the plan and operation of a project. A well concluded security assessment would present documentation drawing any security breaks between a plan design and accepted company security policies. Risk assessment is an inherent part of a broader risk management strategy to help reduce any potential risk-related consequences.[9,10]

IA processes typically begin with the enumeration and classification of information assets to be protected and then proceed to perform a risk assessment of those assets. [8] Risk assessment permits finding vulnerability in the information assets. After the risk assessment is complete, the IA then develops a risk management plan. The plan proposes countermeasures that involve mitigation, elimination, accepting, and transferring the risk and considers prevention, detection, and response to threats.

3- Framework

For the development of countermeasures, IA uses a framework published by a standard such as NIST RMF, Risk It, COBIT, PCI DSS, ISO/IEC 27002, or CERT CSIRT. IA does not seek to eradicate all hazards but to supervise them in the most cost -effective approach. [1, 5]. Later than the risk management map is realized, it is experienced and appraised repeatedly by means of format audit. Information assurance method is known as repetitive, in that the risk assessment and risk management preparation are meant to be occasionally amended and improved based on data collected about their wholeness and efficiency. [2, 8]

4-Benefits

The purpose of information assurance is to reduce risks by creating certain information on which the business achieved appraisal is consistent. The goal of Information assurance is skilled by risk supervision, encryption at rest and in transit, and data integrity. Information assurance guaranteed to present efficient and well-organized ways to defend information systems.

Most important benefits comprise rising responsibility of information security performance; improving success of information security activities; representing compliance with laws, rules, and regulations; and providing experimental inputs for resource allocation conclusions. Information assurance plays a significant part in networked infrastructure, e-commerce, e-business, and e-government. [11]

Information assurance interests in risk management and touches on recommendations for caring for data protection, whether on physical (hard drives, PCs, laptops, and tables) or digital (cloud) systems. AI is alarmed with the business outlook of data. Information assurance hangs out with policymaking and distribution to remain data resources safe. Information assurance recognizes how a group deals with data, the cost of the data, and how revealed that information occurs to be. Information assurance defends data and information methods and incorporates both physical and numerical data.

5-Role in Education

Information assurance was classified as a national main concern in the United States for the security of the important data communications. Yet prior to PDD-63(Presidential Decision Directive 63 attempts to protect infrastructure) being created,

The US government formed the NIETP [12] to, including additional purposes; establish Centers of Academic Excellence in Information Assurance Training and Education. This established an academic infrastructure to reinforce the significant data structure. Creating NIETP exhibits the comprehension of authority superiors in information assurance.

In the United States, the federal government created an ROTC (Reserve Officers Training Corps)-type scholarship program to raise the volume of the information assurance workforce. These Scholarships for Service (SFS) [13] offer maximum tuition and stipends to sponsor both undergraduate and graduate learning in information assurance at centers of excellence schools. Upon graduation, the student is obliged to operate for the federal government for two years.

Such programs are essential to initiate a collection of experts skilled in typical ways to improve and protect governmental information assurance.

## 5. Conclusion.

Information security is all very frequently theorized as somewhat that can be answered with technology. Technology is a significant factor to play; however, it must be the definitive phase in the way out rather than the first..Information assurance is obliged to organize a consistent and complete security approach that can be implemented to secure the company's assets. It is not only for the large global organization; but for all businesses, including the smallest one, to protect and secure the organization's assets. The information security approach requires stakeholders and other mechanisms to be involved, and the information assurance approach involves people and procedures and then supported by technology, which helps an improvement of successful security policies and other strategies to protect the organization.

Information assurance spotlights quality, dependability, and renovation of information. Information security emphasizes an installing security solutions, encryption, policies, and techniques to fix information. IA is not involved with the identifiable technology or devices used to defend information. Somewhat, it is concentrated throughout creating policies and standards. Information security honestly contracts with devices and technologies used to defend information. It's a practical-on methodology that protects data from cyber threats. IA weights managerial hazard control and generally information quality. As a result, information assurance has an expansive extent. Information security worries risk control and arrangement.